\documentclass{emulateapj}

\shorttitle{Single-Wind MHD Model for $\eta$ Car}
\shortauthors{Matt \& Balick}

\slugcomment{Accepted by ApJ}

\begin{document}


\title{Simultaneous Production of Disk and Lobes: \\ A Single-Wind MHD
Model for the $\eta$ Carinae Nebula}

\author{Sean Matt$^{1,}$\altaffilmark{3} and Bruce Balick$^2$}
\affil{$^1$Physics \& Astronomy Department, McMaster University, 
Hamilton ON, Canada L8S 4M1\\ $^2$Astronomy Department, University of 
Washington, Seattle WA 98195\\
matt@physics.mcmaster.ca, balick@astro.washington.edu}

\altaffiltext{3}{CITA National Fellow}


\begin{abstract}

The luminous blue variable $\eta$ Carinae is surrounded by a complex
and highly structured nebula of ejected material.  The best-studied
and axisymmetric components of this outflow consist of bipolar lobes
(the ``homunculus'') and an equatorial ``skirt.''  Recent proper
motion measurements suggest that the skirt was ejected at the same
time as the lobes, contrary to the assumptions of all current
theoretical models for the formation of the nebula (which use the
skirt to collimate stellar winds into lobes).  We present a
magnetohydrodynamic (MHD) stellar wind model that produces an
outflowing disk and bipolar lobes in a single, steady-state wind.  The
basic model consists of a wind from a rotating star with a
rotation-axis-aligned dipole magnetic field. The azimuthal component
of the magnetic field, generated by stellar rotation, compresses the
wind toward the equator and also toward the rotation axis,
simultaneously producing an outflowing disk and jet.  We use numerical
MHD simulations to study the wind for various amounts of stellar
rotation and to show a range of wind morphologies.  In order to
produce wide angle lobes similar to the homunculus (which have roughly
a 30 degree opening angle), a high-speed polar wind (with enhanced
energy density) from the star is also required.  In that case, the
structure of the wind bears a remarkable resemblance to the skirt plus
homunculus morphology of the $\eta$ Car nebulae, and a significant
fraction of the stellar angular momentum is carried away by the
wind. Although the model assumes a steady-state wind (rather than an
eruption) and thermal wind driving (rather than radiation pressure),
the structure of the wind is encouraging.

\end{abstract}


\section{Introduction \label{introduction}}

During the ``great eruption,'' beginning in 1837 and lasting until
about 1858, the luminous blue variable $\eta$ Carinae ejected more
than one solar mass of material \citep[for a review,
see][]{davidsonhumphreys97}. That ejected material has been well
studied \citep[e.g.,][]{davidsonea01} and is highly structured,
consisting of an outflowing equatorial ``skirt'' and bipolar lobes
(the hourglass-shaped ``homunculus'').  The presence of the bipolar
lobes has been explained by some authors \citep*[e.g.,][]{frank3ea95,
dwarkadasbalick98, langer3ea99} with interacting stellar wind (ISW)
models.  In such models, a previously ejected and slowly moving
equatorial flow (presumably related to the skirt) acts as an inertial
barrier that channels a subsequent flow toward the poles, forming the
homunculus \citep[but see also][]{gonzalezea04}.

However, recent proper motion measurements by \citet{morseea01} reveal
that much of the material in the skirt has the same dynamical age as
the lobes.  These observations suggest that the skirt was ejected at
the same time as the lobes (within the uncertainty of a few years) in
1846, during the great eruption, and thus rule out the ISW models.
Clearly, there is a need for a theoretical model that produces bipolar
lobes and an equatorial flow, simultaneously, in a single wind.  As
far as we know, no such model yet exists in the literature.


An alternative to the ISW scenario is shaping of the outflow by some
sort of magnetohydrodynamic (MHD) process.  This explanation has
observational support from \citet{aitkenea95}, who measured the
polarization of mid-infrared emission from the spatially-resolved
homunculus.  Their explanation for the structure and degree of
polarization requires a magnetic field within the homunculus with a
strength in the range of milli-Gauss.  The homunculus is surrounded by
material that was ejected prior to the great eruption, so the magnetic
field in the homunculus cannot be the result of swept-up and
compressed ISM field, and it must have originated from the central
source of the wind \citep{aitkenea95}.

The influence of magnetic fields on stellar winds has been studied by
many authors.  \citet{sakurai85} extended the magnetohydrodynamic
(MHD) wind theory of \citet{weberdavis67} and studied isotropic,
thermally-driven stellar winds, embedded in a radial magnetic
field. Sakurai showed that when the star rotates, the magnetic field
is twisted azimuthally, which results in a collimation of the wind
toward the rotation axis, as well as an increased wind speed
\citep[see also][]{michel69, blandfordpayne82}.  \citet[][hereafter
WS93]{washimishibata93} considered the effect of a dipole magnetic
field on an otherwise isotropic, pressure-driven stellar wind.  They
showed that, when the star rotates rapidly, the azimuthally twisted
dipole produces the same collimation shown by Sakurai, but also
results in an enhancement of the wind along the magnetic equator.  In
other words, the wind from a rapidly rotating star with an aligned
dipolar magnetic field contains both an outflowing disk and a jet.

It seems plausible that the skirt in the $\eta$ Car nebula was
produced by a similar mechanism, as the skirt may correspond to the
outflowing disk in such a wind.  On the other hand, the homunculus
consists of wide angle lobes, and does not match the narrow jet
morphology predicted by the standard, rotating MHD wind theory.  So
the idea of an isotropic wind driving mechanism, plus a rotating
dipole magnetic field, must be modified to produce a wider angle polar
flow.  In this work, we follow the work of WS93 and explore the
possibility that the skirt and homunculus were formed during the great
eruption in a single wind from a rotating star with a dipole magnetic
field.  In order to produce the wide-angle, bipolar lobes of $\eta$
Car, we find that the wind driving must have been more energetic near
the poles than at lower latitudes.  Section \ref{sec_model} contains a
description of the general model and the assumptions therein.

Using numerical, 2.5-dimensional (axisymmetric), MHD simulations, we
follow the acceleration of the stellar wind from the surface of the
star to beyond 100 stellar radii, self-consistently capturing the
interaction between the stellar magnetic field and the wind plasma.
In order to understand each physical process active in the wind and to
demonstrate a range of wind morphologies, we carry out a limited
parameter study.  Section \ref{sec_details} contains a description of
the simulations and the parameters we explored.  Section
\ref{sec_results} contains the results of the parameter study, which
are quite general, and may be applicable to a variety of structured
stellar outflows, particularly those with quadrupolar symmetry, such
as pulsar winds and proto-planetary nebulae.  One of the simulations
produces in a steady-state wind with structures that bear a remarkable
resemblance to the skirt and homunculus morphology of $\eta$ Car.

\section{General Model and Physical Parameters} \label{sec_model}

Axisymmetric outflow nebulae are the result of a coherent but poorly
understood process that probably arises in very close proximity to the
stellar surface.  Thus, one of the ultimate goals of any wind study is
to link observed wind parameters (and morphology) to the conditions at
or very near the source of the wind.  In this paper, we explore the
possibility that the outflow from $\eta$ Car during the great eruption
was produced by a wind from an isolated star.  We hope to constrain
the possible conditions near the surface of $\eta$ Car during the
great eruption.  Knowing these conditions, and how they change in
time, is part of the bigger picture of understanding the internal
structure and evolution of extremely massive stars.

We shall assume that $\eta$ Car had a dynamically important magnetic
field on its surface during outburst.  Magnetic fields are capable of
producing structure in otherwise isotropic flows, particularly in
systems with rotational energy, where the twisting of magnetic field
lines naturally leads to a collimation of the flow along the rotation
axis.  We will employ an MHD formulation of the problem, adopting all
of the assumptions therein.  This assumption is supported, in
principle, by the observations of \citet{aitkenea95}.

In general, any MHD stellar wind is characterized by several key
components: magnetic field geometry, magnetic field strength, stellar
rotation rate and gravity, and the wind driving mechanism.  These
components may have time dependence as well as spatial variations
along the surface of the star.  In addition, they are all likely to be
interrelated.  For example, the rotation rate may influence a stellar
dynamo, which determines the field strength and geometry, which in
turn, may influence the structure and temperature in a corona, etc.
However, since there is currently no comprehensive theory relating all
key components with one another, we treat them as independent
parameters.

Following the work of WS93, we consider pressure (thermal) driving for
the wind, using a polytropic ($\gamma = 1.05$) equation of state, as
appropriate for solar wind studies.  The assumption of thermal driving
is not very appropriate for $\eta$ Car, as radiation pressure is
likely to have been more important during the great eruption
\citep[e.g.,][]{dwarkadasowocki02}.  However, since this is an initial
study, our assumption is for simplicity, and the general results will
be valid for any wind driving mechanism.  Also as in WS93, the
magnetic field anchored in the surface of the star is a rotation axis
aligned dipole with a fixed field strength.  A dipolar magnetic field
is the most reasonable choice, as it is the lowest order multipole
occurring in nature and reflects the symmetries inherent to a rotating
star.  Even the sun's magnetic field is dominated by the dipole
component outside a few solar radii over most of the magnetic cycle
\citep[][]{bravo3ea98}.  Significant progress has been made in recent
years toward understanding the effect of dipole fields on stellar
winds \citep[e.g., WS93;][]{babelmontmerle97a, porter97,
keppensgoedbloed99, keppensgoedbloed00, mattea00, uddoulaowocki02},
and this literature provides a background for the work contained here.

There are four fundamental parameters, one for each of the energies in
the system, which can be written in terms of characteristic speeds.
The thermal energy is a measure of the wind driving (via pressure
gradients) and is represented by the sound speed at the surface of the
star, $c_{\rm s}$.  The gravitational potential energy must be
overcome by the wind and is parameterized by the escape speed from the
surface, $v_{\rm esc}$.  The rotational kinetic energy is represented
by the rotation speed at the equator, $v_{\rm rot}$, and the magnetic
energy by the Alfv\'en speed at the equator, $v_{\rm A}$.  The
conditions in the wind, including the location of various critical
points (where the wind flow speed equals other characteristic speeds),
outflow speed, mass flux, etc., will be completely determined,
self-consistently, from the conditions present on the stellar surface
(i.e., at the base of the wind).

For simplicity, and so that we can first understand the shaping of
individual winds, we only consider steady-state wind solutions.
Below, we present winds with features that resemble both the skirt and
homunculus, but since the $\eta$ Car nebula was ejected in an
outburst, we are only tackling part of the problem.  Specifically, we
assume that our steady-state solution represents the wind from $\eta$
Car during the outburst.  Since the true outburst lasted for a few
years (or decades, at most), the nebula's evolution from outburst
until the present day was likely to have been influenced by previous
and subsequent winds.  Thus, our steady-state models do not address
the conditions responsible for or resulting from the time-dependent
outburst phenomenon as a whole.  However, since there are currently no
other stellar wind models that produce a disk and lobes
simultaneously, we hope that the present work can serve as a starting
point for more complex models, and that it may be more widely
applicable to other classes of structured stellar outflows.

The rotation rate of $\eta$ Car is unknown, so we consider a full
range of rotation rates in our models, from no rotation to a
significant fraction of breakup speed.  Similarly, there are currently
no direct measurements of a magnetic field on the surface of $\eta$
Car.  Furthermore, the possible field strength on the star during the
great eruption, which is relevant here, is completely unconstrained by
observations.  However, we assume that the field was strong enough to
dynamically influence the wind, and we pick a single value for the
field strength.  We choose not to vary the magnetic field, for
simplicity, and since the effect of various dipole field strengths has
been studied by \citet{keppensgoedbloed00} and
\citet{uddoulaowocki02}.  In general, for the magnetic field to be
dynamically important, the magnetic energy density must be comparable
to the kinetic energy density in the wind.  This energy balance gives
a rough lower limit on the magnetic field strength on the stellar
surface of
\begin{eqnarray}
\label{eqn_blimit}
B_{\rm s} \ga 2.8 \times 10^3 \;{\rm G}\;
  {\left({{100 R_\odot}\over{R_*}}\right)}\;  \nonumber \\ \times
  {\left({\dot M_{\rm w}}\over{10^{-1} \; M_\odot/{\rm yr}}\right)}^{1\over 2}\;
  {\left({v_\infty}\over{600 \; {\rm km}/{\rm s}}\right)}^{1\over 2}
%
\end{eqnarray}
where we have assumed that, during the outburst, $\eta$ Car had a
radius of $R_* = 100 R_\odot$ and drove a wind with a mass outflow
rate of $\dot M_{\rm w} = 10^{-1} M_\odot$ yr$^{-1}$ and a speed of
$v_\infty = 600$ km s$^{-1}$.  These assumptions are consistent with
observationally determined values \citep[][]{davidsonhumphreys97}.
For the solar wind, equation \ref{eqn_blimit} gives a minimum field of
0.1 Gauss, though the sun maintains a 1--2 Gauss dipole magnetic field
throughout most of the solar cycle \citep[][]{bravo3ea98}.  By drawing
upon the example of the sun, we choose a dipole field strength of $2.5
\times 10^4$ Gauss, measured at the equator of $\eta$ Car.  This is a
strong field for a giant star, but it is required if magnetic shaping
is to be important.  Note that the wind from $\eta$ Car during
outburst was extremely powerful, compared to other classes of stellar
winds.  If the process that produces an increase in wind energy during
outburst also produces an increase in dynamo activity, it may be
natural to assume that the magnetic field will also exist in a
heightened state during outburst.

We know from the work of WS93 that the effect of a rotating dipolar
field on an otherwise isotropic wind is to, simultaneously, compress
the flow toward the magnetic equator and collimate it toward the
rotation axis.  The resulting outflowing disk superficially resembles
the morphology of the skirt around $\eta$ Car, but the collimated jet
is too narrow to explain the wide-angle, bipolar lobes that comprise
the homunculus.  In order to produce these lobes, the wind driving
could be more energetic near the poles than at lower latitudes---a
similar modification to the rotating dipole wind model was used by
\citet{tanakawashimi02} to reproduce the three-ring structure of SN
1987A.  This enhancement of the polar wind effectively reduces the
dynamical influence of the magnetic field at high latitudes, reducing
the collimation in that region and resulting in a polar flow with a
wide opening angle.  To this end, we present some models with an
increased temperature on the stellar poles, and since we are
considering a thermally driven wind, the higher temperature results in
a more energetic flow from the poles.  There is justification for an
enhanced polar wind from $\eta$ Car.  First, radiatively driven winds
from rapidly-rotating, luminous blue variable stars are expected to be
more energetic in the polar direction \citep{dwarkadasowocki02}.
Second, observations of $\eta$ Car \citep{smithea03, vanboekelea03}
reveal that the present-day stellar wind is significantly enhanced in
the polar direction.  Furthermore, for more general support,
observations of the solar wind show that the kinetic energy and
outflow speed is larger (by a factor of $\sim 2$) above 30 degrees
latitude than at lower latitudes \citep[e.g.,][]{goldsteinea96}.

In summary, we consider steady-state, MHD winds from a star with a
rotation-axis-aligned dipole magnetic field and varying amounts of
rotation.  The wind is thermally driven, and we consider two
possibilities: cases with isotropic wind driving, and cases with an
enhanced polar wind.  The resulting winds (described in \S
\ref{sec_results}) display an encouraging range of morphologies.  The
case with the fastest rotation and with an enhanced polar wind most
resembles the $\eta$ Car nebula.

\section{MHD Simulation Details} \label{sec_details}

Here we present numerical MHD simulations of several models, each with
a different set of stellar surface values of $c_{\rm s}$, $v_{\rm
esc}$, $v_{\rm rot}$, and $v_{\rm A}$.  It is not our intention to
carry out a complete parameter study, which should include variations
of all of these parameters, plus variations of the magnetic field
geometry, and wind driving anisotropy.  Instead, our aim is to
determine what surface conditions may be required to reproduce the
sort of structures seen in the $\eta$ Car outburst (i.e., the skirt
and homunculus).

The cases we ran were chosen to illustrate, in a step-by-step manner,
the individual effect of each physical process, leading up to a
``final'' model that resembles the $\eta$ Car nebula.  In each case,
the parameters are chosen and held fixed on the stellar surface, and
the simulations run until the entire computational domain is filled
with a steady wind emanating from the stellar boundary.  This method
gives the steady-state solution of the flow within the computational
grid and determined solely by the conditions held fixed on the stellar
surface boundary.

     \subsection{Numerical Method}

We use the 2.5-dimensional MHD code of \citet{mattea02}, which we
describe briefly here.  The reader will find further details of the
code in \citet[][and also in \citealp*{goodson3ea97,
matt02}]{mattea02}, which solves the ideal (non-resistive) MHD
equations using a two-step Lax-Wendroff, finite difference scheme
\citep{richtmyermorton67} in cylindrical $(r, \phi, z)$ geometry.  The
formulation of the equations allows for a polytropic equation of state
(we use $\gamma$ = 1.05), includes a source term in the momentum and
energy equations for point-source gravity, and makes the explicit
assumption of axisymmetry ($\partial / \partial \phi$ = 0 for all
quantities).  The fundamental plasma quantities in the equations are
the vector magnetic field $\mbox{\boldmath $B$}$, mass density $\rho$,
vector momentum density $\rho \mbox{\boldmath $v$}$ (where
$\mbox{\boldmath $v$}$ is the velocity), and energy density $e = 0.5
\rho v^2 + P / (\gamma - 1)$, where $P$ is the thermal pressure.  Due
to the geometry and symmetry in the system, it is often useful to
decompose the vectors into the azimuthal and poloidal components.  The
azimuthal component is that in the cylindrical $\phi$ direction and
denoted by a subscript ``$\phi$,'' and the poloidal component is that
in the $r$-$z$ plane denoted by a subscript ``p'' (e.g.,
$\mbox{\boldmath $v$} = v_{\rm p} \hat {\rm p} + v_\phi \hat \phi$ and
$v_{\rm p} \hat {\rm p} = v_r \hat r + v_z \hat z$).

The computational domain consists of four nested, grids (or ``boxes'')
in the cylindrical $r$-$z$ plane.  Each box contains $401 \times 400$
gridpoints (in $r$ and $z$, respectively) with constant grid spacing.
The boxes are nested concentrically, so that the inner box represents
the smallest domain at the highest resolution.  The next outer box
represents twice the domain size with half the spatial resolution (an
so on for other boxes).  With four boxes, the total computational
domain is eight times larger than that of the innermost grid, but
requiring only four times the computational expense.  This
computational efficiency allowed us to run the several cases necessary
for the current study.  A circular boundary, centered on the origin
and with a radius of 30 gridpoints, represents the surface of the
star.  Assuming $R_* = 100 R_\odot$ for $\eta$ Car during outburst,
the innermost grid then has a spatial resolution of roughly $3.33
R_\odot$, and a domain size of $13.3 R_*$.  Similarly, the outermost
(fourth) box has a resolution and domain size of $26.7 R_\odot$ and
$107 R_*$, respectively.

     \subsection{Boundary and Initial Conditions}

We use standard outflow conditions on the box boundaries, appropriate
for wind studies.  Namely, along the outermost boundary in $r$ and $z$
in the fourth box, a continuous boundary condition (in which the
spatial derivative across the boundary is zero for all quantities)
allows the stellar wind to flow out of the computational domain
undisturbed.  In all grids, we enforce reflection symmetry across the
equatorial ($z = 0$) plane.  Along the rotation axis ($r = 0$), we
require the azimuthal and radial components of magnetic field and
velocity to be zero, and all other quantities are equal to their value
at $r = d r$ (where $d r$ is the grid spacing).

The surface of the star is represented by a spherical inner boundary,
centered at the origin ($r = z = 0$).  This boundary is the source of
the wind that is to fill the computational domain, and so the
conditions here entirely determine the solution of the system.
Following the wind from the very surface of a star and
self-consistently capturing the interaction between the wind plasma
and a magnetic field that is anchored into the rotating stellar
surface is a complex problem.  In a real star, the properties of the
wind (mass flux, velocity, etc.)  are determined by the balance of
forces resulting from (e.g.)\ thermal pressure gradients, gravity,
centrifugal forces, and through interaction with the poloidal magnetic
field (which distorts when pushed and also pushes back) and the
azimuthal magnetic field---which is generated by a twisting of the
poloidal field as outflowing plasma tries to conserve its angular
momentum.

In order to capture these interactions within the framework of our
second-order finite difference scheme, we employ a four-layer boundary
for the star, on which the various parameters are set as follows: For
all gridpoints such that $R \le 34.5$ (where $R = (r^2 + z^2)^{1/2}$
in units of the grid spacing in the innermost box), the poloidal
velocity is forced to be parallel with the poloidal magnetic field
($v_{\rm p} \parallel B_{\rm p}$).  Where $R \le 33.5$, $\rho$ and $P$
are held constant (in time) at their initial values.  For $R \le
32.5$, $v_{\rm p}$ is held at zero, while $v_\phi$ is held at
corotation with the star.  For $R \le 31.5$, $B_{\rm p}$ field is held
at its initial, dipolar value, while $B_\phi$ is set so that there is
no poloidal electric current at that layer (which gives it a
dependence on the conditions in the next outer layer, $31.5 < R \le
32.5$).  We consider the spherical location $R = 30$ to be the
surface of the star, as this is where all quantities are held fixed
and is thus the absolute base of the wind.

These boundary conditions capture the behavior of a wind accelerated
thermally from the surface of a rotating magnetized star, as follows:
There is a layer on the stellar boundary ($R > 32.5$) outside of which
the velocity not fixed, but is allowed to vary in time.  In this way,
the wind speed and direction is determined by the code in response to
all of the forces.  By holding $P$ fixed at its initial value for all
$R \le 33.5$, we constrain the pressure gradient force (thermal
driving) behind the wind to be constant.  Also, holding the density
fixed at $R \le 33.5$ allows the region from where the wind flows to
be instantly replenished with plasma.  Thus, the base of the wind
maintains a constant temperature and density, regardless of how fast
or slow the wind flows away from that region.  The existence of a
layer in which $v_{\rm p} = 0$ and $B_{\rm p}$ can evolve (namely, at
$31.5 < R \le 32.5$) allows $B_{\rm p}$ (and $v_{\rm p}$) to reach a
value that is self-consistently determined by the balance of magnetic
and inertial forces (and is also necessary to maintain a negligible
$\nabla \cdot \mbox{\boldmath $B$}$).  We set the poloidal velocity
parallel to the poloidal magnetic field for the next two outer layers
(which ensures a smoother transition from the region of pure dipole
field and zero velocity to a that with a perturbed field and outflow)
for reasons of numerical stability.  Setting $B_\phi$ so that the
poloidal electric current is zero inside some radius ensures that the
field is completely anchored in a rotating conductor (the surface of
the star).  This way, $B_\phi$ evolves appropriately outside the
anchored layer according to the interaction with the wind plasma.

We have extensively tested these conditions for a wide range of all
system parameters, and have been able to reproduce the work of other
authors (e.g., the results of \citealp{keppensgoedbloed99} and WS93).
The advantage of our boundary conditions is in their
versatility---they capture the expected behavior for a wide range of
conditions.  Of particular note, they produce the correct hydrodynamic
solution for the case with $\mbox{\boldmath $B$} = 0$, and give the
same solution for a similar case with a weak (dynamically
insignificant) dipole magnetic field---in which dipole field lines are
stripped open to a radial, split monopole configuration by the
spherical wind (see, e.g., \citealp{mattea00}).  In cases with a
strong dipole, closed magnetic loops near the equator overcome thermal
and centrifugal forces to ``shut off'' the flow, resulting in a
magnetically closed region with $v_{\rm p} = 0$ and that corotates
with the star (i.e., self-consistently forming a ``dead zone''; see
\citealp{keppensgoedbloed00}).

For the simulations with enhanced polar winds (see \S
\ref{sub_parmspace}), the pressure at high latitudes (above
$74^\circ$) on the star is held fixed at twice the pressure at the
same spherical radius but at lower latitudes.  Formally, for all
regions on the star such that $R \le 33.5$ and $z/r \ge 3.5$, the
pressure is doubled.  The resulting factor of two pressure
discontinuity across $74^\circ$ latitude on the star is only slightly
larger than the typical radial pressure gradients near the star, in
which, at the resolution of our simulations, the variation in pressure
between gridpoints is typically a factor of 1.3.  Furthermore, the
code has been tested, and behaves well, under shock conditions, in
which discontinuities can be much larger than a factor of two.  Thus,
the enhanced polar wind does not present a difficulty for the
numerical method.

Ultimately, the conditions in the steady-state wind that emanates from
the stellar surface depends only on the parameters held fixed on the
star.  However, an initial state for the system must be specified on
the entire simulation grid, so we initialized the grid with the
spherically symmetric wind solution of \citet{parker58}.  As discussed
above, the pressure gradient and density given by this initial
\citet{parker58} solution is held fixed for all time on the star,
though the velocity is allowed to vary to respond to the rotation and
magnetic field of the star.  This approach is similar to the work of
\citet{keppensgoedbloed99}.  For the cases with an enhanced polar
wind, the pressure is doubled on the polar region of the star, so the
fixed pressure gradient is also double.

     \subsection{Parameter Space Explored \label{sub_parmspace}}

Our approach is to illustrate, in a logical progression of cases, the
individual effects of each additional parameter.  To this end, we ran
the 12 simulation models listed in table \ref{tab_parms}.  The cases
named ``ISO,'' have isotropic thermal wind driving, in which the
pressure from the initial \citet{parker58} solution is held fixed at
all latitudes on the star.  The number in the last part of the name
corresponds to the rotation period in days, assuming the fiducial
radius for $\eta$ Car during outburst used in equation
\ref{eqn_blimit}.  The rotation period is also listed in the second
column of table \ref{tab_parms}.  The cases named ``HTC'' are those in
which the thermal wind driving is enhanced on a ``high temperature
cap'' on the star.  They are identical to the corresponding ISO cases,
except that for all regions on the star above a latitude of
$74^\circ$, the pressure ($P$) is set and held at double the value of
the ISO case.  Thus, the star has a polar cap with twice the
temperature (and twice the pressure gradient and thermal energy) as
for lower latitudes, which results in a more energetic flow from the
polar region.

\begin{deluxetable}{lrcc}
\tablewidth{8.3cm}
\tablecaption{Simulation Parameters\tablenotemark{a}$\;$ \label{tab_parms}}
\tablehead{
\colhead{Case} &
\colhead{$T$} &
\colhead{$v_{\rm rot} \over v_{\rm esc}$\tablenotemark{c}} &
\colhead{$v_{\rm A} \over v_{\rm esc}$\tablenotemark{c}} \\
\colhead{Name\tablenotemark{b}} &
\colhead{(days)} &
\colhead{} &
\colhead{}
}

\startdata

ISO$^0\infty$     & $\infty$ & 0     & 0  \\
ISO$\infty$       & $\infty$ & 0     & 0.35 \\
ISO2000           & 2000     & 0.004 & 0.35 \\
ISO200            & 200      & 0.041 & 0.35 \\
ISO100            & 100      & 0.082 & 0.35 \\
ISO50             & 50       & 0.165 & 0.35 \\

HTC$^0\infty$     & $\infty$ & 0     & 0    \\
HTC$\infty$       & $\infty$ & 0     & 0.35 \\
HTC2000           & 2000     & 0.004 & 0.35 \\
HTC200            & 200      & 0.041 & 0.35 \\
HTC100            & 100      & 0.082 & 0.35 \\
HTC50             & 50       & 0.165 & 0.35 \\

\enddata

\tablenotetext{a}{The value of $c_{\rm s}/v_{\rm esc}$ is the same in
all models and equal to 0.223; see text.}

\tablenotetext{b}{The ``ISO'' models have isotropic wind driving,
while the ``HTC'' models include a high temperature polar region on
the star.}

\tablenotetext{c}{Given is the value at the equator and surface of the
star.}

\end{deluxetable}

In all models, the mass and radius of the star are assumed to be $100
M_\odot$ and $100 R_\odot$, respectively, which corresponds to an
escape speed of $v_{\rm esc} = 617$ km s$^{-1}$.  The sound speed ($=
\sqrt{\gamma P/\rho}$) on the surface of the entire star in the ISO
models and below 74$^\circ$ latitude for the HTC models is $c_{\rm s}
= 137$ km s$^{-1}$ (and so $c_{\rm s} = 194$ km s$^{-1}$ in the polar
region of the HTC models).  The values of $c_{\rm s}$ and $v_{\rm
esc}$ determine the solution to the velocity at all radii in a
(non-rotating, non-magnetic) hydrodynamic wind.  We chose this value
of $c_{\rm s}$ so that the outflow speed reaches several hundred km
s$^{-1}$, far from the star, and we set our density normalization
(which is the same for all models) such that the mass outflow rate is
around $10^{-1} M_\odot$ yr$^{-1}$, appropriate for $\eta$ Car during
outburst.  The magnetic field, when nonzero, is always a
rotation-axis-aligned dipole with a strength of $B_* = 2.5 \times
10^4$ Gauss on the equator of the star (and twice that on the pole).
With our density normalization, this gives an Alfv\'en speed ($=
B/\sqrt{4 \pi \rho}$ in cgs units) on the surface of the star at the
equator of $v_{\rm A} = 218$ km s$^{-1}$.  We explore a full range of
stellar rotation up to a rotation of 23\% of breakup speed.  Thus the
equatorial rotation speed on the star for different models ranges from
$v_{\rm rot} = 0$ to 102 km s$^{-1}$.

The ratios of $v_{\rm rot}/v_{\rm esc}$ and $v_{\rm A}/v_{\rm esc}$ at
the equator and surface of the star are listed in the third and fourth
columns of table \ref{tab_parms} for each model.  The ratio of $c_{\rm
s}/v_{\rm esc}$ is 0.223 for all models (except on the polar region of
the HTC models, where $c_{\rm s}/v_{\rm esc} = 0.315$).  In order to
establish a baseline for the ISO cases, we first ran a purely
hydrodynamical case ($B = 0$) with no rotation, called ISO$^0\infty$.
Then, to show the effect of a $2.5 \times 10^4$ G dipolar field, we
ran the ISO$\infty$ case, also with no rotation.  Next, to illustrate
the twisting of magnetic fields, but before the twisting is
dynamically important, we ran the ISO2000 case, in which the rotation
is so slow that the solution is virtually identical to the ISO$\infty$
case (see \S \ref{sec_results}).  The effects of dynamically
significant rotation are demonstrated with the ISO200, ISO100, and
ISO50 cases.  Finally, we ran an HTC model corresponding to each of
the ISO models, so that we could see the effect of the anisotropic
thermal wind driving for each set of parameters.

The results of these simulations (presented in \S \ref{sec_results})
are scalable to any star with the same values of $v_{\rm rot}/v_{\rm
esc}$, $v_{\rm A}/v_{\rm esc}$, and $c_{\rm s}/v_{\rm esc}$.  Note
that the rotation period (and any physical times) should then scale
with $R_* / v$, where $v$ represents any of the characteristic speeds
in the system (e.g., $v_{\rm esc}$).  In section \ref{sub_global} we
calculate the outflow rates of mass ($\dot M_{\rm w}$), angular
momentum ($\dot J_{\rm w}$), and energy ($\dot E_{\rm w}$), which can
also be scaled.  The mass outflow rate scales with $v R_*^2$ times the
density normalization, $\dot J_{\rm w} \propto \dot M_{\rm w} v R_*$,
and $\dot E_{\rm w} \propto \dot M_{\rm w} v^2$.  Finally, the
magnetic field strength ($B$) scales with $v$ times the square root of
the density normalization (so that $B \propto \sqrt{\dot M_{\rm w} v}
/ R_*$).  As an example, the sun has the same $v_{\rm esc}$ and
roughly the same coronal sound speed and $v_{\rm A}$ as our models.
Using $\dot M_{\rm w} \sim 10^{-14} M_\odot$ yr$^{-1}$, for the sun,
scaling down $\dot M_{\rm w}$ and $R_*$ in our models corresponds to a
dipole field strength of $\sim 1$ Gauss, which is appropriate
\citep{bravo3ea98}.  Thus the ISO2000 model is similar to the sun,
though with a rotation period of 20 days.  Since the solar rotation
does not have a significant effect on the solar wind (see, e.g.,
WS93), we expect the ISO2000 case to display similar global properties
as the solar wind (though the solar wind is further modified by
anisotropic coronal heating, high order magnetic fields, etc.).

     \subsection{Accuracy of the Numerical Solution \label{sub_accuracy}}

After the start of the simulations, a flow emanates from the surface
of the star.  In all cases, that flow accelerates to a speed faster
than all information carrying waves (i.e., slow, Alfv\'en, and fast
magnetosonic waves) within the computational domain, usually within
the first or second grid.  To obtain a numerical steady-state
solution, we ran each case for $\sim 2$ years of physical time.  Since
it takes only $\sim 0.6$ yr for the wind, traveling with a typical
velocity of 400 km s$^{-1}$, to cross our largest simulation box,
material that leaves the surface of the star at the beginning of the
simulation has traveled well beyond the entire domain by the end.  A
measure of the fractional rate of change of each of the quantities
$\rho$, $P$, $v_r$, $v_\phi$, $v_z$, $B_r$, $B_\phi$, and $B_z$ at
every gridpoint quantifies the steadiness of the numerical solution.
In the largest box, for each case, the average gridpoint was changing
by less than 1\% per wind crossing time by the end of the simulations.
Since the computations require roughly 16,000 time-steps per crossing
time, this corresponds to a fractional change of less than $10^{-6}$
per time-step.  As another simple test of the steadiness and
conservation of our numerical solutions, we calculated the outflow
rates of $\dot M_{\rm w}$, $\dot J_{\rm w}$, and $\dot E_{\rm w}$ (see
\S \ref{sub_global}) at all radii within the computational domain and
found that these global quantities are constant (i.e., no variation
with radius), within 3\%, for all cases.  Thus, our solutions are
sufficiently steady for the present study.

One general physical test of numerical MHD solutions is to quantify
the divergence of $\mbox{\boldmath $B$}$, which should always and
everywhere be zero.  The initial state of the system has $\nabla \cdot
\mbox{\boldmath $B$} = 0$, and the code maintains this physical
requirement to second order in space and time.  However, numerical
errors do lead to non-zero values of $\nabla \cdot \mbox{\boldmath
$B$}$, and we deal with these errors in two ways.  First, the code
solves the momentum equation using a non-conservative formulation of
the magnetic force that prevents numerical $\nabla \cdot
\mbox{\boldmath $B$}$ errors from producing unphysical forces
\citep{brackbillbarnes80}.  Second, we check that these numerical
errors do not contribute significantly to the structure of the
magnetic field, by computing the ratio of $\nabla \cdot
\mbox{\boldmath $B$}$ to $|\nabla B|$---that is, the magnitude of the
gradient of the magnitude of $\mbox{\boldmath $B$}$---at every
gridpoint.  This ratio determines the maximum possible contribution of
non-zero $\nabla \cdot \mbox{\boldmath $B$}$ to any spatial change in
the magnetic field.  For the vast majority of gridpoints, in all
cases, this ratio is always much less than 1\%.  Near the stellar
boundary, this can be larger, though never exceeding 5\% in all
magnetic cases.  These $\nabla \cdot \mbox{\boldmath $B$}$ errors are
acceptable.

For a more comprehensive check on the precision of our solutions, we
follow the error analysis of \citet{ustyugovaea99}, who list five
integrals of motion $K$, $\Lambda$, $\Omega$, $S$, and $E$ (their
eqs.\ 1--5), corresponding to the conservation of mass, angular
momentum, helicity, entropy, and energy, respectively.  Under the
conditions of ideal, axisymmetric, steady-state MHD, these integrals
should be constant along a given magnetic field line, and thus serve
as a stringent test of our numerical solution.  In every magnetic
model, we calculated $K$, $\Lambda$, $\Omega$, $S$, and $E$ along two
reference field lines: one that connects to the stellar surface at
$65^\circ$ latitude, and one that connects to $80^\circ$.  In this
way, we sampled two latitudes in the flow, one of which is within the
hot polar cap of the HTC models.  In all cases, all five integral
functions are conserved to within 5\%, along both reference field
lines.

For an even more stringent test, we compared the absolute value of the
conserved quantity $\Omega$ to the known value at the stellar surface,
which equals the constant angular rotation rate of the star.
Following the error analysis of \citet{keppensgoedbloed00}, we
calculated $\Omega$ for all gridpoints in our domain, for all of the
magnetic models.  Over 90\% of the domain, the error in $\Omega$
ranges from 2--7\%, for all cases, except one.  This exception is the
ISO50 case, in which this error is 12\%.  A larger error in $\Omega$
exists in all cases over an area of less than 10\% of the domain that
covers a narrow region near the equator with an opening angle of less
than $10^\circ$, as seen by the star.  This error occurs in the
portion of the flow that travels along the lowest latitude open field
lines near the dead zone (see \S \ref{sec_results}).  In that portion
of the flow, the maximum error in $\Omega$ ranges from 20--40\% in all
cases, except one.  This exception is the HTC2000 case, in which the
maximum error in this portion of the flow reaches 80\%.  These errors
are comparable to those of \citet{keppensgoedbloed00}, who found
errors in $\Omega$ reaching 40\% in the same portion of the flow as
where our largest errors occur.

\section{Simulation Results} \label{sec_results}

For the cases with nonzero magnetic field, the wind is significantly
altered from the purely hydrodynamical solution.  The behavior of the
dipole magnetic field in a wind and it's effect on the thermally
driven flow is discussed by many authors \citep[see
especially][]{keppensgoedbloed00}.  In particular, the dipole field
lines at high latitudes are nearly radial, and thus wind can flow
along them with little impedance.  Further, they reach to distances
far from the star, where the field becomes very weak.  As the dipole
field energy density falls off with $R^{-6}$, and the kinetic energy
density in the wind falls off more slowly than $R^{-2}$ (for an
accelerating wind) these high latitude field lines will open up as
they are dominated by and carried out in the wind.  On the other hand,
at lower magnetic latitudes, the dipole field lines are more
perpendicular to the flow, and they do not reach as far away from the
star.  If the field is strong enough to counteract the outward thermal
pressure forces, the flow will be shut off.  This produces a region of
closed field lines around the magnetic equator, a ``dead zone,'' that
corotates with the star and from which no wind can flow (in the
absence of magnetic diffusion).

\begin{figure*}
\epsscale{1.}
\plotone{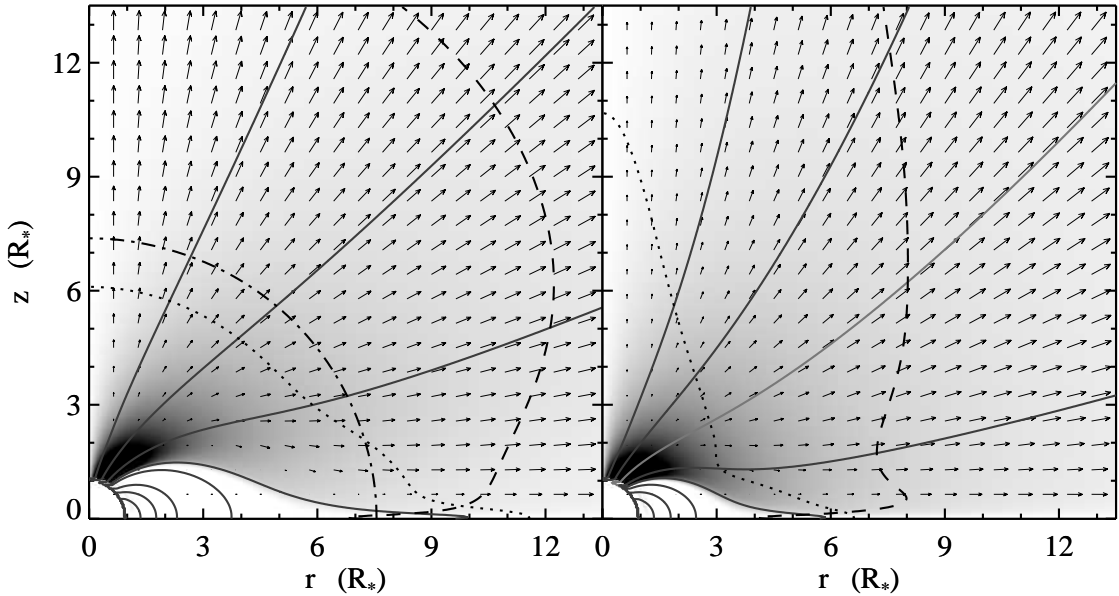}

\caption{Greyscale of $B_\phi$ (white is zero, black is strong) with
selected magnetic field lines overdrawn in grey for two cases where
the star has a constant temperature across its surface.  The dashed
line indicates the radial Alfv\'en surface, and the dotted line is the
sonic surface.  The left panel contains data for the very weakly
rotating case ($T = 2000$ days), while the right panel corresponds to
faster rotation ($T = 100$ days). The dash-dotted line in the left
panel is the location of the sonic surface for a simulation without
rotation, and without a magnetic field.  The light grey field line in
the right panel intersects the stellar surface at $65^\circ$ and
indicates the path $s$ followed in Figure \ref{fig_forces}.
\label{fig_bphi_nocap}}

\end{figure*}

If the star rotates, wind material tries to conserve its angular
momentum as it travels outward so that its angular velocity decreases
with distance from the rotation axis.  When wind plasma is connected
to the stellar surface by magnetic fields, the difference in angular
rotation rate between the star and wind twists the field azimuthally,
generating a $B_\phi$ component to the field \citep[see,
e.g.,][]{parker58}.  If the stellar rotation is fast enough, magnetic
forces associated with $B_\phi$ significantly influence the flow
(e.g., WS93).  These effects will be further discussed below.

     \subsection{Wind Acceleration and Shaping \label{sub_shaping}}

The left panel of Figure \ref{fig_bphi_nocap} illustrates the
conditions in the steady-state wind for the ISO2000 case.  Shown is a
greyscale image of $B_\phi$ (black represents 57 Gauss, and white is
zero), the poloidal magnetic field lines (solid lines), and poloidal
velocity vectors (arrows).  The data is from the innermost (highest
resolution) computational box.  The star is at the lower left and has
a radius of $R_*$.  Poloidal magnetic field lines are drawn so that
they originate at roughly evenly spaced intervals in latitude on the
star (their spacing is not proportional to field strength).  We have
also drawn a field line that marks the transition between closed and
open field regions (i.e., the line that encloses the dead zone).  Note
that, under steady-state, ideal MHD conditions (as we have here),
magnetic field lines are also in the same direction as plasma flow
streamlines.  The rotation is so slow in the ISO2000 case that the
resulting wind is identical (within a few percent) to the ISO$\infty$
case, except that $B_\phi$ is exactly zero for ISO$\infty$.  Thus, the
left panel of Figure \ref{fig_bphi_nocap} simply demonstrates the
effects of a $2.5 \times 10^4$ G dipole magnetic field on an otherwise
completely isotropic flow.  The results here are comparable to other
similar studies \citep[e.g.,][]{keppensgoedbloed00}, and they form the
base to our understanding of the effects of additional processes in
the wind (i.e., rotation and enhanced polar winds).

In the left panel of Figure \ref{fig_bphi_nocap}, the dead zone covers
a significant fraction of the stellar surface near the equator and
extends beyond 5 $R_*$ above the star.  Thus, all of the wind from the
star originates from the open field region at high latitudes.  There
is flow at low latitudes, outside several stellar radii, but (as
indicated by the field lines and velocity vectors) that flow
originates from the polar region of the star.  In other words, the
wind from the poles diverges very rapidly near the star and tends more
toward radial divergence far from the star.  The flow actually
converges toward the equator within roughly $8 R_*$, as the open field
lines curve around the dead zone.  The dotted line in the left panel
of Figure \ref{fig_bphi_nocap} marks the location of the sonic
surface, where the wind velocity equals the local sound speed.  For
reference, the dash-dotted line marks the sonic surface for the
ISO$^0\infty$ case, in which there is no magnetic field, and the wind
is completely isotropic.  The effect of a dipole field is to shift the
sonic surface further from the star near the equator.  This is because
the convergent flow on the equator leads to a shallower radial
pressure gradient, and the wind is accelerated more slowly there.
Conversely, the sonic surface is moved closer to the star near the
pole because the flow diverges faster than radial divergence, making
the pressure gradient steeper.  The dashed line marks the location of
the poloidal Alfv\'en surface, where the poloidal wind velocity equals
the local poloidal Alfv\'en speed (calculated from the $B_{\rm p}$
component of the field only).  The Alfv\'en surface is furthest from
the star along the pole, since the magnetic field is strongest there.
Due to the opening of field lines in the wind, the poloidal magnetic
field strength goes to zero on the equator (across which there is a
direction reversal of the field, supported by a current sheet),
leading to a ``cusp'' in the Alfv\'en surface there.

The greyscale of $B_\phi$ in the left panel of Figure
\ref{fig_bphi_nocap} shows that the dead zone is corotating with the
star, since $B_\phi \approx 0$ there.  However, in the open field
region, the field lines are ``trailing'' the star as it rotates
beneath the expanding wind, and so $B_\phi \neq 0$.  Note that
$B_\phi$ is zero along the rotation axis because the rotation speed
goes to zero at 90$^\circ$ latitude. $B_\phi$ is also zero everywhere
along the equator because the radial component of the magnetic field
goes to zero there, for a dipole, and it is the radial component of
the magnetic field that is twisted to generate $B_\phi$.  Thus, for a
twisted dipole field, $B_\phi$ has a maximum value at mid latitudes
\citep[][]{washimi90}.  There is a magnetic force associated with
gradients in $B_\phi$ (proportional to $- \nabla B_\phi^2$), which
point toward the rotation axis \citep[as studied by,
e.g.,][]{sakurai85} and also toward the equator (as pointed out by
WS93).  For the ISO2000 case, these magnetic forces are too small to
have a significant effect, and the wind is accelerated
hydrodynamically, though modified by the greater (less) than radial
divergence near the polar (equatorial) region, imposed by the poloidal
magnetic field.  

For the cases with more rapid rotation, however, the forces associated
with $B_\phi$ become important, and act both perpendicular to and
parallel to the poloidal field to modify the flow.  This is evident in
the right panel of Figure \ref{fig_bphi_nocap}, which illustrates the
steady-state flow for the ISO100 case.  The quantities shown in the
right panel are the same as for the left panel, except that the
greyscale image of $B_\phi$ is scaled so that black corresponds to 860
Gauss, and white is zero.  The only difference between the two cases
is that the star in the ISO100 case rotates 20 times faster than for
ISO2000, so a stronger $B_\phi$ is generated that influences the wind.
A comparison between the two panels of Figure \ref{fig_bphi_nocap}
reveals this influence.  First, the dead zone is significantly smaller
for ISO100 because the magnetic pressure gradient in $B_\phi$, acting
perpendicular to the poloidal field, is strong enough to ``squash''
the closed field region.  In addition, the field lines in the ISO100
case exhibit greater than radial divergence from the polar region of
the star (as in the ISO2000 case) but then undergo a change in
curvature outside a few $R_*$ and begin to diverge more slowly than
radially near the rotation axis.  This gives rise to an asymptotic
collimation of the flow, and is caused by the magnetic pressure
gradients in $B_\phi$ that are perpendicular to the poloidal field and
directed toward the axis.  This collimation results in shallower
pressure gradients along the pole, leading to slower acceleration of
the wind there, and the sonic point is shifted outward relative to the
ISO2000 case.  This effect (and due to the more slowly diverging
poloidal magnetic field) also moves the Alfv\'en surface further out
on the poles.  The most important difference between the left and
right panels of Figure \ref{fig_bphi_nocap}---that a rotating dipole
field simultaneously compresses the flow on the magnetic equator and
collimates it along the rotation axis---was first shown by WS93 (e.g.,
compare with their fig.\ 1).

\begin{figure}
\epsscale{1.1}
\plotone{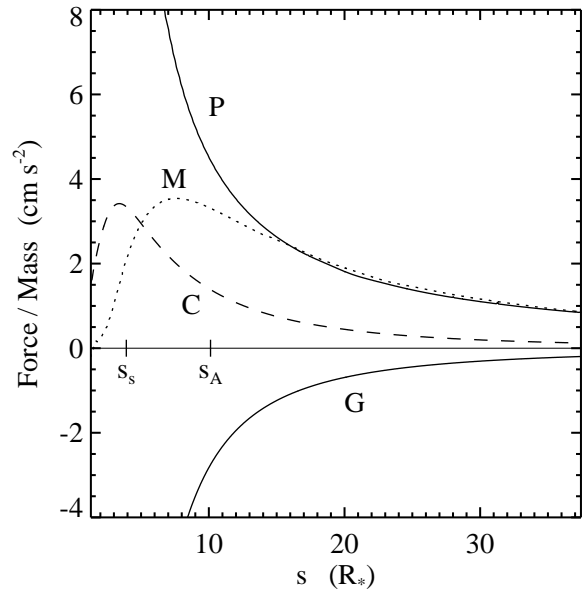}

\caption{Forces parallel to the reference field line indicated by the
light grey line in right panel of Figure \ref{fig_bphi_nocap}, as a
function of position $s$ along the field line.  Lines labeled P, G, C,
and M represent the pressure gradient, gravitational, centrifugal, and
magnetic forces, respectively (eqs.\ \ref{eqn_fp}--\ref{eqn_fm}).  The
locations $s_{\rm s}$ and $s_{\rm A}$ indicate the sonic and
Alfv\'enic critical points, respectively.
\label{fig_forces}}

\end{figure}

\begin{figure*}
\epsscale{1.0}
\plotone{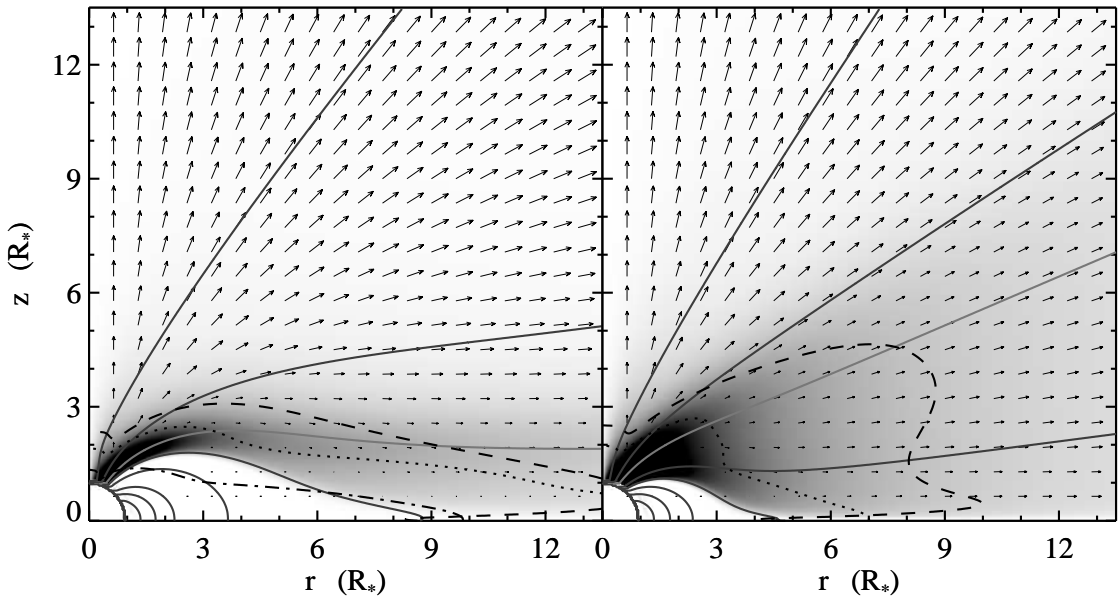}

\caption{Greyscale of $B_\phi$ (white is zero, black is strong) with
selected magnetic field lines overdrawn in grey for two cases where
the star has a high temperature polar cap.  The light grey field line
in each panel intersects the stellar surface at 74$^\circ$, the
latitude of the temperature discontinuity on the star.  The dashed
line indicates the radial Alfv\'en surface, and the dotted line is the
sonic surface.  The left panel contains data for the very weakly
rotating case ($T = 2000$ days), while the right panel corresponds to
faster rotation ($T = 100$ days). The dash-dotted line in the left
panel is the location of the sonic surface for a simulation without
rotation, and without a magnetic field.
\label{fig_bphi}}

\end{figure*}

Fast rotation also affects the flow in the direction parallel to the
poloidal magnetic field.  Figure \ref{fig_forces} shows these forces
along a reference magnetic field line that connects to the star at
$65^\circ$ latitude, for the ISO100 case.  This field line is
indicated by the light grey line in the right panel of Figure
\ref{fig_bphi_nocap}, and the data in Figure \ref{fig_forces} is from
the second computational box.  The forces per mass along a field line
are given by \citep[e.g.,][]{ustyugovaea99}:
\begin{eqnarray}
\label{eqn_fp}
f_{\rm P} &=& - {1 \over \rho} {\partial P \over \partial s},  \\
\label{eqn_fg}
f_{\rm G} &=& - {G M_* \over R^2} \hat R \cdot \hat s,  \\
\label{eqn_fc}
f_{\rm C} &=& v_\phi r \; \hat r \cdot \hat s, \\
\label{eqn_fm}
f_{\rm M} &=& 
  -{1 \over 8 \pi \rho r^2} {\partial (r B_\phi)^2 \over \partial s},
\end{eqnarray}
where the subscripts P, G, C, and M correspond to the pressure
gradient, gravitational, centrifugal, and magnetic forces,
respectively.  Here, $s$, is the distance along the poloidal field,
$G$ is the gravitational constant, and $M_*$ is the mass of the star.
Also note that we have used both the cylindrical ($r$) and spherical
($R$) radial coordinate, where appropriate.  Figure \ref{fig_forces}
shows these forces in units of cm s$^{-2}$ (for $\eta$ Car parameters,
where the surface gravity is 270 cm s$^{-1}$), as a function of
position along $s$, where $1 R_*$ corresponds to the stellar surface.
It is evident that the pressure gradient force along the reference
field line dominates near the star, but the centrifugal and magnetic
forces are significant.  We find that the ratio of $f_{\rm C} / f_{\rm
P}$ reaches a maximum value of $\sim 30$\% at $\sim 10 R_*$ along $s$.
Consequently, material at mid latitudes is partially
magnetocentrifugally accelerated, leading to a faster outflow.  As a
result, the sonic and Alfv\'enic surfaces are closer to the star than
for the ISO2000 case (see Fig.\ \ref{fig_bphi_nocap}).

The left and right panels of Figure \ref{fig_bphi} show the
steady-state winds of the HTC2000 and HTC100, respectively, in the
same format as Figure \ref{fig_bphi_nocap} (the grey-scaling for
$B_\phi$ is also the same for each respective panel).  The two cases
shown in Figure \ref{fig_bphi} are identical to those in Figure
\ref{fig_bphi_nocap} except that the star now has a high temperature
polar region.  In each panel of Figure \ref{fig_bphi}, there is a
light grey field line that is anchored on the stellar surface at
$74^\circ$ latitude.  Since that is the latitude of the edge of the
high temperature region, and since the field line is also a
streamline, it marks the division between material that originates
from the hot polar cap and material that flows from lower latitudes.

A comparison of the left panels of Figures \ref{fig_bphi_nocap} and
\ref{fig_bphi} reveals the effect of the hot polar cap on the wind
from a slowly rotating star with a dipole field.  The increase in
thermal energy on the pole more quickly accelerates the wind, which
moves the sonic and Alfv\'en surface much closer to the star near the
pole.  However, since the hot polar wind is not in pressure
equilibrium with the wind at lower latitudes, it expands rapidly
toward the equator as it flows outward (more than for ISO2000), until
there is a meridional balance in pressure between the high and low
latitude flows.  Far from the star, the flow in a large range of
latitudes originates from the hot polar cap (further discussed in \S
\ref{sub_morph}).  The convergence of flow toward the equator reduces
radial pressure gradients there, which moves the sonic and Alfv\'en
surface outward (relative to the ISO2000 case).  The azimuthal
magnetic field exhibits the same qualitative behavior in both the ISO
and HTC cases.

A comparison between the left and right panels of Figure
\ref{fig_bphi} reveals the effect of significant rotation in the HTC
models.  As with the ISO cases, centrifugal flinging accelerates the
flow more rapidly in the equatorial direction, moving the location of
the sonic and Alfv\'en surface inward.  Also, the dead zone is
somewhat squashed by the magnetic forces (associated with $B_\phi$)
directed toward the equator.  However, in contrast to the ISO cases,
the dynamical effects of the rotation in the HTC case are not as
important at high latitudes, due to the stronger pressure gradient
there.  Along a field line connected to the HTC100 star at $80^\circ$
latitude, for example, $f_{\rm C} / f_{\rm P}$ reaches a maximum value
of less than 1\%, compared to a ratio of greater than 9\% along a
field line connected to the star at $65^\circ$.  This difference
between the high and low latitude flow is evident in the right panel
of Figure \ref{fig_bphi}, as the lowest latitude open field line is
bending slightly poleward (collimating), while the higher latitude
field lines are not.  The high latitude flow is is inherently more
energetic, so it would require a stronger field and/or faster rotation
than the lower latitude flow, to be significantly altered.
Consequently, the effect of significant rotation on the HTC models is
to ``clear out'' the wind at mid latitudes as it is compressed toward
the equator and also against the side of the polar flow.  Note that
the direction of the light grey line points to higher latitudes in the
HTC100 case than in the HTC2000 case, indicating that, when the star
spins faster, the energetic polar flow is confined to a narrower
opening angle.

     \subsection{Global Outflow Properties \label{sub_global}}

\begin{figure*}
\epsscale{1.0}
\plotone{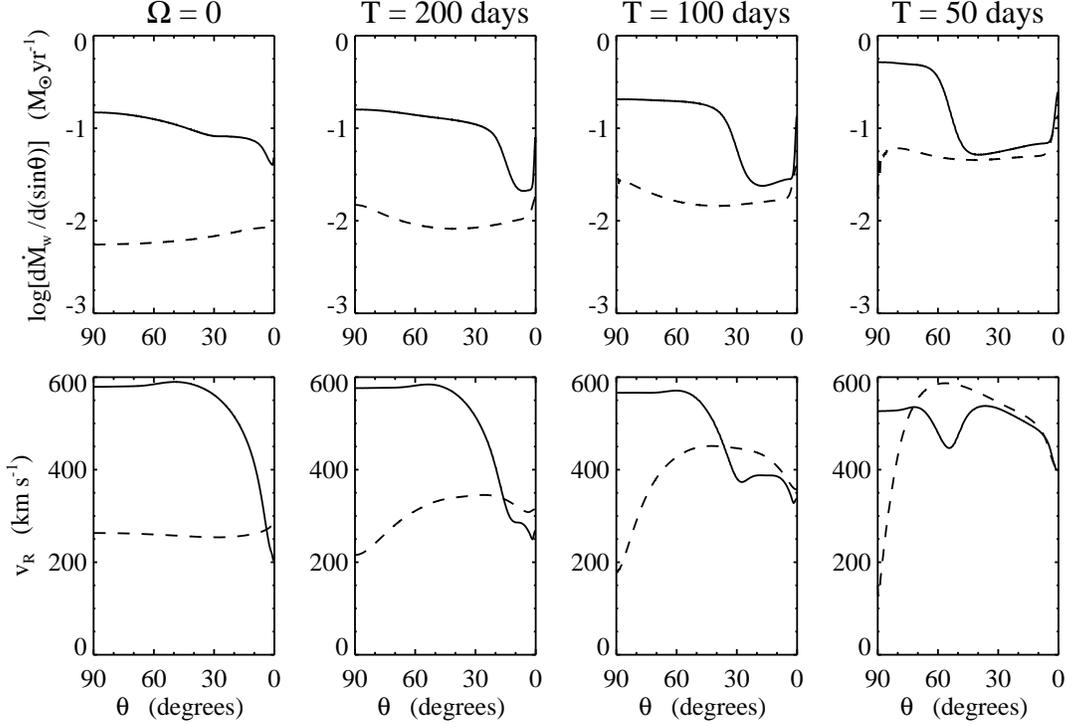}

\caption{The mass outflow rate, $\partial\dot M/ \partial(\sin\theta)$, (top
panels) and radial velocity of the wind (bottom panels) as a function
of latitude, $\theta$, for cases with (from left to right) no
rotation, $T = 200$, $T = 100$, and $T = 50$ days.  The dashed lines
represent cases with spherical thermal wind driving, while the solid
lines are for the cases with a high temperature polar wind.  All
panels are calculated at a distance of $40 R_*$ from the star.
\label{fig_angplots}}

\end{figure*}

The presence of a dipole field and the rotation of the magnetized star
influences the global outflow rates of mass ($\dot M_{\rm w}$),
angular momentum ($\dot J_{\rm w}$), and energy ($\dot E_{\rm w}$).
We calculate these quantities from the simulation data by integrating
the fluxes over an enclosed spherical surface at a radius $R$.
Thus, 
\begin{eqnarray}
\label{eqn_mdot}
\dot M_{\rm w} &=& 
  4 \pi R^2 \int_{0}^{1} \rho v_R \; d(\sin \theta),  \\
\label{eqn_jdot}
\dot J_{\rm w} &=& 
  4 \pi R^2 \int_0^{1} \rho v_R \; \Lambda \; d(\sin \theta),  \\
\label{eqn_edot}
\dot E_{\rm w} &=& 
  4 \pi R^2 \int_0^{1} \rho v_R \; E^\prime \; d(\sin \theta), 
\end{eqnarray}
where $\theta$ is the latitude, $v_R$ is the velocity component in the
spherical radial direction ($v_R = \mbox{\boldmath $v$} \cdot \hat
R$), and
\begin{eqnarray}
\label{eqn_lambda}
\Lambda = v_\phi r -  {B_\phi B_{\rm p} r \over 4 \pi \rho v_{\rm p}},  \\
\label{eqn_e}
E^\prime = {v_{\rm p}^2 + v_\phi^2 \over 2}
  + {\gamma \over \gamma - 1} {P \over \rho}
  - {G M_* \over R}     \nonumber \\
  + {B_\phi^2 \over 4 \pi \rho} 
  - {v_\phi B_\phi B_{\rm p} \over 4 \pi \rho v_{\rm p}}.
\end{eqnarray}
Here, $\Lambda$ is the integral of motion discussed in section
\ref{sub_accuracy} that gives the angular momentum per mass carried in
the wind, and $E^\prime = E + \Lambda \Omega$ gives the specific
energy carried by the wind.  The first three terms in equation
\ref{eqn_e} represent the kinetic ($E_{\rm K}$), thermal, and
gravitational energy, while the last two terms together represent the
magnetic energy ($E_{\rm M} =$ Poynting flux divided by $\rho v_R$).
Note that the equations \ref{eqn_mdot}--\ref{eqn_edot} have been
multiplied by a factor of two to take both hemispheres into account.

The top row of panels of Figure \ref{fig_angplots} shows the
differential mass loss rate, $\partial \dot M_{\rm w} / \partial (\sin
\theta) = 2 \pi R^2 \rho v_R$ (``mass flux'').  The bottom row of
panels shows the radial velocity $v_R$.  Both the mass flux and $v_R$
are plotted versus $\theta$, at a constant spherical radius of $40
R_*$ (which equals 300 gridpoints in the third computational box), for
several models.  From left to right, the panels for both mass flux and
$v_R$ show cases with increasing rotation rate.  In each panel, the
dashed line is for the ISO case, while the solid line represents the
HTC case.  Note that for the ISO2000 and HTC2000 cases, which are
featured in the left panels of Figures \ref{fig_bphi_nocap} and
\ref{fig_bphi}, the rotation is so slow that the mass flux and $v_R$
for those models are the same (within a few percent) as the data
plotted in the left panels of Figure \ref{fig_angplots} for the
ISO$\infty$ and HTC$\infty$ cases.  Effectively, then, Figure
\ref{fig_angplots} displays data from all 10 of the models with
non-zero magnetic field.


The dashed lines in the left panels of Figure \ref{fig_angplots}
reveal that steady-state wind in the non-rotating ISO$\infty$ case is
not completely isotropic.  For reasons discussed in section
\ref{sub_shaping}, the flow is densest near the equator \citep[see
also][and \S \ref{sub_morph}]{mattea00}.  This is the effect of the
dipole field on an otherwise isotropic wind.  For the HTC models, the
flow is generally faster at higher latitudes, and the mass flux is
higher than for the ISO cases, due to the more energetic wind driving
in the HTC cases.

\begin{deluxetable*}{lccccccc}
\tablewidth{16cm}
\tablecaption{Simulation Results \label{tab_results}}
\tablehead{
\colhead{Case} &
\colhead{$\dot M_{\rm w}$} &
\colhead{$\dot J_{\rm w}$} &
\colhead{$\dot E_{\rm w}$} &
\colhead{$\dot E_{\rm K}$\tablenotemark{a}} &
\colhead{$\dot E_{\rm M}$\tablenotemark{a}} &
\colhead{$v_{\rm max} \over v_{\rm min}$\tablenotemark{b}} &
\colhead{$\rho_{\rm max} \over \rho_{\rm min}$\tablenotemark{b}} \\
\colhead{Name} &
\colhead{($10^{-1} M_\odot$/yr)} &
\colhead{($10^{45}$ erg)} &
\colhead{($10^{39}$ erg/s)} &
\colhead{($10^{39}$ erg/s)} &
\colhead{($10^{39}$ erg/s)} &
\colhead{} &
\colhead{}
}

\startdata

ISO$^0\infty$     & 0.15 & \nodata & 2.2 & 0.44 & \nodata & 1.0 & 1.0 \\
ISO$\infty$       & 0.14 & \nodata & 2.1 & 0.42 & \nodata & 1.1 & 1.5 \\
ISO2000           & 0.14 & 0.16    & 2.1 & 0.42 & 0.0029  & 1.1 & 1.5 \\
ISO200            & 0.19 & 1.6     & 3.4 & 0.86 & 0.22    & 1.6 & 2.9 \\
ISO100            & 0.34 & 3.7     & 7.3 & 2.4  & 0.84    & 2.6 & 5.0 \\
ISO50             & 1.1  & 11      & 28  & 11   & 3.9     & 4.2 & 4.3 \\

HTC$^0\infty$     & 2.6  & \nodata & 86  & 27   & \nodata & 2.0 & 1.9 \\
HTC$\infty$       & 1.8  & \nodata & 61  & 20   & \nodata & 2.9 & 1.7 \\
HTC2000           & 1.8  & 0.13    & 61  & 20   & 0.0043  & 2.9 & 1.7 \\
HTC200            & 1.9  & 2.0     & 62  & 20   & 0.48    & 2.3 & 4.0 \\
HTC100            & 2.0  & 4.9     & 66  & 20   & 2.0     & 1.7 & 6.5 \\
HTC50             & 2.7  & 13      & 88  & 26   & 7.9     & 1.3 & 10.2 \\



\enddata

\tablenotetext{a}{Given is the value at a radius of $80 R_*$.}
\tablenotetext{b}{Given is the value at a radius of $40 R_*$.}

\end{deluxetable*}

The integrated outflow rates $\dot M_{\rm w}$, $\dot J_{\rm w}$, and
$\dot E_{\rm w}$ are listed in the second, third, and fourth columns
of table \ref{tab_results}, for each case.  Since $\dot E_{\rm w}$ is
dominated by the thermal energy within the computational domain, due
to our chosen value of $\gamma$ near unity, we list the individual
components of $\dot E_{\rm K}$ and $\dot E_{\rm M}$ in the fifth and
sixth columns of table \ref{tab_results}.  These are calculated with
equation \ref{eqn_edot} but using only the kinetic or magnetic terms
of equation \ref{eqn_e}.  The table lists these values at a spherical
radius of $80 R_*$ (which equals 300 gridpoints in the fourth
computational box), where the gravitational energy is negligible.  If
one scales $\dot M_{\rm w}$ and $R_*$ down to solar values for the
ISO2000 case (as described in \S \ref{sub_parmspace}), this case
predicts $\dot J_{\rm w} \approx 1.1 \times 10^{30}$ erg for the solar
wind.  This value is comparable to the predictions of $\dot J_{\rm w}
\sim 10^{30}$ erg from the solar models of \citet{keppensgoedbloed00},
and this further verifies our numerical solution.

It is evident from table \ref{tab_results} that $\dot M_{\rm w}$ and
$\dot E_{\rm w}$ are slightly less in the ISO$\infty$ case than for
ISO$^0\infty$ (and similarly for HTC$\infty$ and HTC$^0\infty$).  This
is because the existence of the dead zone ``shuts off'' part of the
flow.  However $\dot M_{\rm w}$ and $\dot E_{\rm w}$ are not simply
reduced by the amount proportional to the fractional area covered by
the dead zone on the star, since a reduced wind flux near the equator
is counteracted, somewhat, by an increased flow from the poles, driven
by the faster than radial divergence of the wind there.  In order to
quantify the anisotropy of the wind, we use a crude measure given by
the ratio of the maximum value of $v_R$ to the minimum $v_R$ found in
the wind at a spherical radius of $40 R_*$.  This ratio is listed in
the seventh column of table \ref{tab_results} for all cases, while the
eighth column contains a similar ratio for the anisotropy in the mass
density.

From the data in table \ref{tab_results} and plotted in Figure
\ref{fig_angplots}, it is evident that, as the stellar rotation rate
increases, $\dot M_{\rm w}$, $\dot J_{\rm w}$, $\dot E_{\rm w}$, and
the anisotropy of the flow increases.  The effect of strong azimuthal
magnetic field is to clear out material from mid latitudes, where it
has a maximum strength, as it both compresses material toward the
equator, forming an outflowing disk, and collimates material toward
the pole.  For the ISO models, a jet appears as an increase in mass
flux along the rotation axis, as evident in the top row Figure
\ref{fig_angplots}.  For the HTC models, the opening angle of the
energetic flow from the pole of the star is decreased---an outflowing
``lobe'' appears---as rotation increases.  These features are
discussed further in section \ref{sub_morph}.  Note that for the ISO50
and HTC50 cases, the mass fluxes and $v_R$ are similar at latitudes
less than roughly $40^\circ$.

Also evident in Figure \ref{fig_angplots} is that, in each case, there
is a general anti-correlation between the mass flux and the wind
velocity.  That is, where the mass flux is high relative to the
non-rotating case, the velocity is relatively low.  This is because
the convergent flow that increases the local mass flux also makes the
thermal pressure gradients shallower, leading to a slower acceleration
of the flow.  However, in spite of the general decrease in $v_R$ near
the rotation axis and equator, we find that the linear momentum and
kinetic energy flux is generally the largest where the mass flux is
largest.

An exception to the mass flux and wind velocity anti-correlation
occurs for the ISO50 case (dashed line in the upper right panel of
Fig.\ \ref{fig_angplots}), where the mass flux is enhanced near the
rotation axis (as in other cases), but decrease to a minimum value on
the axis, resulting in a ``hollow'' jet structure.  This behavior is
similar to cold, magnetocentrifugally driven winds, in which the inner
part of the jet is generally hollow \citep[e.g.,][]{ouyedpudritz97}.
The flux in the center of this jet is small because the magnetic field
cannot centrifugally fling material along the exact rotation axis,
where the rotation speed goes to zero.  In our models, we find that,
along a field line that connects at $\theta = 65^\circ$ on the star,
$f_{\rm C} / f_{\rm P}$ reaches a maximum value of 10\%, 30\%, and
70\% for the ISO200, ISO100, and ISO50 cases, respectively.  So the
sequence of ISO models is showing a transition from a purely thermal
wind, to a thermo-centrifugal wind (see WS93).  This is also evident
in table \ref{tab_results}, where the ratio of $\dot E_{\rm M} / \dot
E_{\rm w}$ increases with the rotation rate.


     \subsection{Morphology: Disks, Jets, and Bipolar Lobes \label{sub_morph}}

\begin{figure*}
\epsscale{1.0}
\plotone{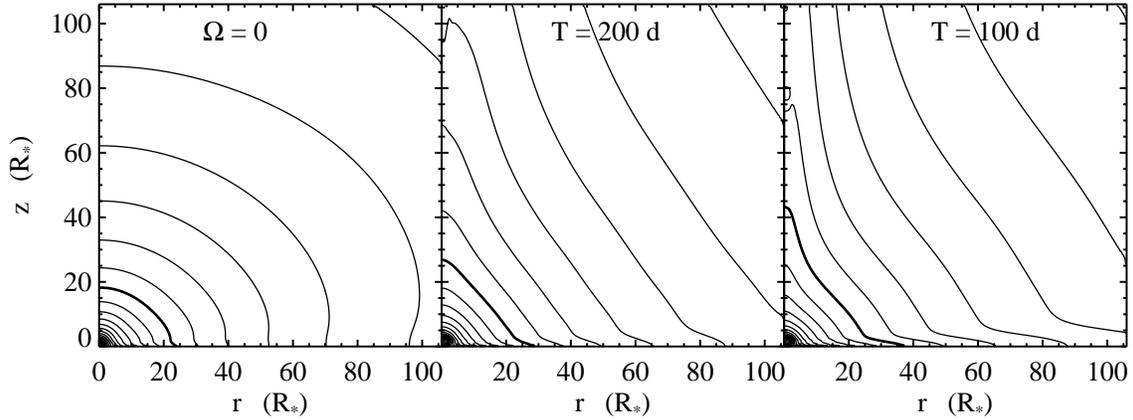}

\caption{Logarithmic density contours in the $r$-$z$ plane for cases
with (from left to right) no rotation, $T$ = 200, and $T$ = 100 days
and where the star has a constant temperature across its surface.
Shown is data from the outermost computational grid, and the star is
at the origin.  The thickest contour line corresponds to a density of
$1.7 \times 10^{-13}$ g cm$^{-3}$, and the spacing between each
contour is a factor of two.
\label{fig_denscont_nocap}}

\end{figure*}

\begin{figure*}
\plotone{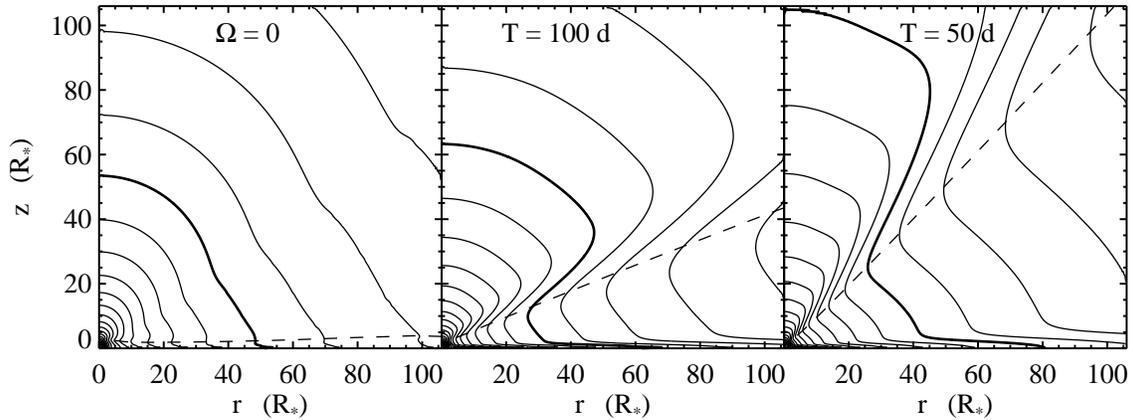}

\caption{Same as Figure \ref{fig_denscont_nocap}, but for cases with
(from left to right) no rotation, $T$ = 100, and $T$ = 50 days and
where the star has a high temperature polar cap.  The dashed line in
each panel shows the magnetic field line that intersects the stellar
surface at 74$^\circ$, the latitude of the temperature discontinuity
on the star. 
\label{fig_denscont}}

\end{figure*}

The outermost (fourth) simulation box reaches beyond $100 R_*$, and
the conditions in the steady-state wind there are indicative of the
shape of the flow very much further from the star.  The three panels
of Figure \ref{fig_denscont_nocap} show logarithmic density contours
from the fourth box for three ISO cases with different stellar
rotation rates (from left to right: ISO$\infty$, ISO200, ISO100).  For
reference, the thickest contour line in each panel corresponds to a
density of $1.7 \times 10^{-13}$ g cm$^{-3}$.  The left panel shows
the effect of the dipole field alone---the wind is denser near the
equator than along the pole.  For stars that rotate quickly (middle
and right panel), notice the formation of the jet along the rotation
($z$) axis and the disk along the equator.  Also, compare the middle
panel with figure 2 of WS93.

Figure \ref{fig_denscont} is the same as Figure
\ref{fig_denscont_nocap} but for the HTC$\infty$, HTC100, and HTC50
models (from left to right).  The left panel shows the conditions in
the wind from a star with both a dipole magnetic field and an
energetic polar wind.  The density is enhanced along the equator, due
to the dipole field, but also along the poles from the increased
thermal driving there.  The dashed line in each panel follows the
magnetic field line that originates at a latitude of $74^\circ$ on the
stellar surface (as the light grey lines in Fig.\ \ref{fig_bphi}).  So
the material that exists at a higher latitude than the dashed line
originates from the high temperature region on the star.  In the left
panel (where the star does not rotate), only material at very low
latitudes originates from the cooler region on the star.  When stellar
rotation becomes important (middle and right panels), the mid
latitudes are cleared out as magnetic pressure gradients compress
material toward the equator and against ``walls'' of the polar flow.
Instead of a narrow jet, wide-angle, bipolar lobes are formed.
Ultimately, the opening angle of the lobes is determined by the
balance between the thermal pressure in polar flow and magnetic
pressure (associated with $B_\phi$) in the low-latitude flow.

\begin{figure}
\epsscale{1.2} \plotone{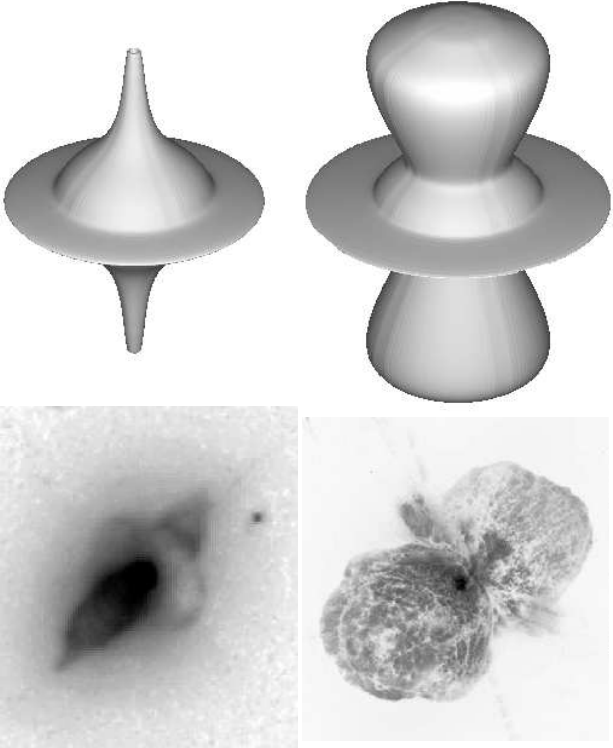}

\caption{Isodensity surface in the steady-state wind for two different
models (top), each with a stellar rotation period of 50 days, compared
to two HST/WFPC2 images of nebulae (bottom).  Top left: density
surface at $5.2 \times 10^{-14}$ g cm$^{-3}$ for a case with isotropic
thermal wind driving.  Top right: density surface at $1.7 \times
10^{-13}$ g cm$^{-3}$ for a case with an enhanced polar wind.  These
are projections with the rotational symmetry axis tilted $30^\circ$
from the plane of the sky.  Bottom left: proto-planetary nebula IRAS
17106--3046 \citep{kwok3ea00}. Bottom right: $\eta$ Carinae
\citep{morseea98}.
\label{fig_isodens}}

\end{figure}

For the HTC50 case (right panel of Fig.\ \ref{fig_denscont}), the
density contours superficially resemble $\eta$ Car's homunculus.  The
morphology of the flow is more clearly seen in the projected,
3-dimensional, isodensity surfaces plotted in the top row of Figure
\ref{fig_isodens}.  On the top left is an isodensity surface in the
steady-state wind of the ISO50 case, and the top right is from the
HTC50 case.  The surface on the top right corresponds to the same
density as the thick line in the right panel of Figure
\ref{fig_denscont}, which serves as a size reference for the 3-D
projections of Figure \ref{fig_isodens}.  To generate these images, we
have exploited the symmetries in the system by rotating the 2-D
density contour from the simulation about the rotation axis and
reflecting it across the equator.  Notice that the disk is similar for
the two cases, but the polar flows are drastically different.  For
comparison, we have also included HST images of two nebulae on the
bottom row of the Figure.  The bottom left is the proto-planetary
nebulae IRAS 17106--3046 \citep*[from][]{kwok3ea00}, and the bottom
right is $\eta$ Car \citep[from][]{morseea98}.  Both nebulae have
their symmetry axes along a line from bottom left to top right, both
exhibit equatorial structures, and both are oriented such that the
axis pointing toward bottom left of the image is tilted out of the
plane by (very roughly) 30$^\circ$.

The top right panel of Figure \ref{fig_denscont} bears a remarkable
resemblance to the skirt and homunculus morphology of $\eta$ Car
\citep[bottom right; also see][]{davidsonea01}.  However, the
comparison is only superficial because we have considered only the
conditions in a steady-state wind, while the lobes and skirt of $\eta$
Car were ejected in an outburst.  Also, the lobes of $\eta$ Car are
hollow \citep{davidsonea01}.  It is not clear from this preliminary
study what sorts of morphologies would result, if the wind were, for
example, ``turned on'' for a few years (simulating an outburst) and
then ``turned off'' again afterwards and left to drift for 150 years.
Such a study is left for future work.

\section{Summary and Discussion \label{discussion}}

We have studied a thermally-driven wind from a star with a
rotation-axis-aligned dipole magnetic field and varying amounts of
rotation.  We have considered both isotropic thermal driving and that
which is enhanced on the pole of the star.  Using MHD simulations, we
determined the 2.5-dimensional (axisymmetric), steady-state wind
solution.  Our parameter study resulted in winds with a wide range of
shapes.  Winds are slightly enhanced along the magnetic equator by the
presence of a dipole magnetic field.  When the star rotates, the
poloidal magnetic field is twisted azimuthally, and magnetic pressure
gradients associated with $B_\phi$ direct material away from mid
latitudes---toward the rotation axis and toward the magnetic equator.
For cases with isotropic thermal driving, fast stellar rotation
results in an outflowing disk and jet morphology.  For cases with an
enhanced polar wind, there is an outflowing disk and wide-angle
bipolar lobes.  As far as we know, no physical model for producing
such a wind (with a simultaneous disk and lobe morphology) has yet
been demonstrated in the literature.

     \subsection{Implications for $\eta$ Car}

The disk and lobe morphology of our steady-state wind solutions
resembles the skirt and homunculus morphology of the $\eta$ Car
nebula, though the latter was formed in a single eruptive outburst and
is much more complex than the flows of our study.  The opening angle
of the lobes in our model HTC50 most resembles that of the homunculus.
In addition, the total energy outflow rate in that model is $\sim 9
\times 10^{40}$ erg s$^{-1}$ (table \ref{tab_results}), which is
comparable to the observed luminosity of $\eta$ Car during outburst of
$\sim 8 \times 10^{40}$ erg s$^{-1}$ \citep{davidsonhumphreys97}.  For
a comparison between the HTC50 model and the $\eta$ Car nebula to be
appropriate, $\eta$ Car needs to have had a roughly axis-aligned
dipole magnetic field of the order of $2.5 \times 10^4$ Gauss during
the great eruption in the mid 1800's.  It also needed to have a wind
driving mechanism that produced a flow that was inherently more
energetic on the poles and needed to be rotating at a significant
fraction of breakup speed \citep[both of which are also predicted
by][]{dwarkadasowocki02}.  Assuming, for now, that $\eta$ Car's wind
resembled the HTC50 model during the great eruption, our model implies
several things.

First, we find that $B_\phi \approx 30$ G at $40 R_*$ and at mid
latitudes.  Assuming $r^{-1}$ dilution of the field and a size of
$37500$ AU for the $\eta$ Car nebula \citep{davidsonhumphreys97}, this
predicts a $\sim 15$ mG (azimuthally directed) field in the
present-day homunculus. This is larger than the $\sim$mG fields
detected by \citet{aitkenea95}, but considering the uncertainties in
our assumed parameters, and that the magnetic field can be dissipated
by other processes, this comparison should not be stringent.  Another
unique prediction of magnetic wind theory is that the entire flow
should be rotating \citep[as verified for protostellar
jets;][]{bacciottiea02}.  The HTC50 model has a rotational speed of
$\sim 30$ km s$^{-1}$ at $40 R_*$ in the disk.  Assuming angular
momentum conservation in the wind from that point outward, this
predicts a rotation speed of $\sim 20$ m s$^{-1}$ in the skirt at
25000 AU, which would be difficult to detect observationally.

The rotation of the outflow extracts angular momentum from the star,
and we find a loss rate of $\approx 1.3 \times 10^{46}$ erg (table
\ref{tab_results}) during outburst.  The angular momentum of $\eta$
Car, for the assumed parameters, is $J_* = k^2 M_* \Omega_* R_*^2
\approx 1.4 \times 10^{55} k^2$ erg s (where $k$ is the normalized
radius of gyration).  Assuming $k^2 \sim 0.1$--$0.2$, the total amount
of angular momentum extracted by the wind during the outburst would be
comparable to $J_*$ in only 3.5--7 years!  The reason for such a large
$\dot J_{\rm w}$ is mainly because of the large mass of the
homunculus, which is $\ga 1$\% of the mass of the star.  Particularly,
if this much mass is shaped magnetocentrifugally, as we have
considered here, $\dot J_{\rm w}$ will necessary be large.  This
result suggests either that $\eta$ Car can have only one or two such
outbursts during its lifetime, or that the star must gain angular
momentum from an external source.  Note that $\eta$ Car may have a
binary companion \citep[see, e.g.,][]{smithea04}, which could likely
further influence the structure of the wind and possibly spin up
$\eta$ Car \citep{smithea03}, though a detailed calculation remains to
be done.



     \subsection{General Application}

The range of morphologies exhibited in our parameter study may be
applicable (with appropriate modifications) to a wide range of
observed outflow nebulae.  A general conclusion of our study is that,
for any magnetic rotator wind in which collimation occurs, if the
source has an aligned dipolar field, the formation of an outflowing
disk is inevitable.  So any outflow nebulae that exhibits quadrupolar
symmetry (i.e., outflowing disk plus jets or disk plus lobes) may be
explained with winds similar to those presented here.  For example,
the general torus plus jet morphology of the crab pulsar x-ray nebula
\citep[e.g.,][]{hesterea02} could be explained by a wind accelerated
in a rotating dipole field, though relativistic effects and other
pulsar physics should be considered.  In addition, proto-planetary
nebulae all show axisymmetries, and the overwhelming majority of
optically visible proto-planetary nebulae have disks seen in
silhouette (e.g., see IRAS 17106--3046 in the lower left panel of
Figure \ref{fig_isodens}).  A mechanism that generates disks and lobes
together may turn out to be important.  The general model presented
here is also appealing because of the relatively small number of
parameters, and its ability to link observed parameters to the
conditions very near the source of the wind.

It is important to note, however, that many astrophysical nebulae of
various types are evidently shaped by processes that are inherently
time-dependent, either through the interaction of multiple winds or in
outburst-type events.  Time-dependence adds many additional parameters
to any outflow model, making parameter studies challenging, but
greatly enriching the range of possible outcomes.  Future work with
magnetocentrifugal stellar winds should include time-dependent
effects.







\acknowledgements

We would like to thank the anonymous referee, whose suggestions led to
significant improvements in the paper.  This research was supported by
NASA grant GO 9050, awarded from the Space Telescope Science Institute
and by the National Science and Engineering Research Council of
Canada, McMaster University, and the Canadian Institute for
Theoretical Astrophysics.















\end{document}